# Structural Stability and Defect Energetics of ZnO from Diffusion Quantum Monte Carlo

Juan A. Santana,[1] Jaron T. Krogel,[1] Jeongnim Kim,[1] Paul R. C. Kent,[2,3] Fernando A. Reboredo[1,a]

[1] Materials Science and Technology Division, Oak Ridge National Laboratory, Oak Ridge, Tennessee 37831, USA

[2] Center for Nanophase Materials Sciences, Oak Ridge National Laboratory, Oak Ridge, Tennessee 37831, USA

[3] Computer Science and Mathematics Division, Oak Ridge National Laboratory, Oak Ridge, Tennessee 37831, USA

**ABSTRACT:** We have applied the many-body *ab-initio* diffusion quantum Monte Carlo (DMC) method to study Zn and ZnO crystals under pressure, and the energetics of the oxygen vacancy, zinc interstitial and hydrogen impurities in ZnO. We show that DMC is an accurate and practical method that can be used to characterize multiple properties of materials that are challenging for density functional theory approximations. DMC agrees with experimental measurements to within 0.3 eV, including the band-gap of ZnO, the ionization potential of O and Zn, and the atomization energy of $O_2$, ZnO dimer, and wurtzite ZnO. DMC predicts the oxygen vacancy as a deep donor with a formation energy of 5.0(2) eV under O-rich conditions and thermodynamic transition levels located between 1.8 and 2.5 eV from the valence band maximum. Our DMC results indicate that the concentration of zinc interstitial and hydrogen impurities in ZnO should be low under *n*-type, and Zn- and H-rich conditions because these defects have formation energies above 1.4 eV under these conditions. Comparison of DMC and hybrid functionals shows that these DFT approximations can be parameterized to yield a general correct qualitative description of ZnO. However, the formation energy of defects in ZnO evaluated with DMC and hybrid functionals can differ by more than 0.5 eV.

[a] Electronic mail: reboredofa@ornl.gov

# I. INTRODUCTION

The development of electronic-structure methods and better computational infrastructures have allowed the building of large open-access materials databases.[1–5] As a result, high-throughput (HT) computation has emerged as a promising tool to help in the design of materials. In fact, HT computational approaches have already been applied to design materials: including electronic, magnetic and multiferroic materials, and materials for catalysis and renewable energy.[6] However, the success of HT approaches for materials design largely depends on the ability of electronic structure methods to describe new and existing materials with the required accuracy. Kohn-Sham density functional theory (DFT)[7] is the standard electronic structure method employed in HT computation of materials. The form of the exchange-correlation functional, the most critical quantity in DFT, is, however, not provided from the theory. DFT methods are therefore based on approximated exchange-correlation functionals, e.g., the local density (LDA), generalized gradient (GGA) and hybrid functional approximations.[7] In turn, these approximations are partially based on many-body quantum Monte Carlo (QMC) calculations of the homogeneous electron gas or the spherical jellium.[8] For an extensive range of materials, these approximations are sufficiently accurate but, unfortunately, they are not accurate enough for many others, particularly materials based on transition-metal oxides. LDA and GGA do not account properly for exchange-correlation effects in transition-metal oxides, and hybrid functionals depend on empirical parameters, undermining their predictive power.

The increasing data being generated by HT computation based on DFT approximations are likely to be unreliable for many important technological materials. A universal and accurate *ab-initio* approach is needed to provide a reliable validation. A method capable of accurately describing the properties of transition-metal oxides without empirical inputs will provide an *ab-initio* description of these complex systems, with the potential to yield new physical insights and drive the discovery of new materials. A natural solution to overcome the intrinsic limitations of the DFT approximations is to directly apply QMC methods to study real materials instead of models. QMC methods, such as the fixed-node diffusion QMC (DMC) method,[9,10] are based on many-body wave functions, capturing all electron dynamics on an equal footing across different systems and offering accurate, parameter-free calculations of materials.

QMC calculations are very expensive computationally, but recent developments in algorithms[11–15] have allowed the application of these methods to study real materials. For instance, DMC has recently been applied to study different properties of transition-metal oxides,[10,16–18] $MgSiO_3$ perovskite,[19] boron-nitride,[20] and point defects in MgO,[21,22] Al,[23] Si[24–26] and diamond.[27] Other QMC methods have also recently been applied to study transition-metal oxides.[28,29] However, DMC or similar QMC methods have not been used to study defects in transition-metal oxides.

Defects play a central role in the physical properties of transition-metal oxides. For instance, oxygen vacancies can turn an oxide from an insulator to a conductor and even a superconductor, or alter the magnetic structure from ferromagnetic to antiferromagnetic.[30] Accurate characterization of defects in transition-metal oxides is crucial to control their functionalities. Computational modeling has become an invaluable tool to help characterize and identify defects because the experimental characterization is rather complex and often indirect. However, the vast majority of calculations of defects in transition-metal oxides are based on DFT approximations.[31–33] DFT results, though, are only semiquantitative as the incorrect description of exchange-correlation effects results in severely underestimated band-gap in transition-metal oxides, introducing significant uncertainties in DFT calculations of defects.

ZnO is one of the best known examples of the limitations of DFT to study defects in transition-metal oxides. ZnO is a wide band-gap semiconductor that crystallizes with the wurtzite structure under ambient conditions.[34] It can also exist with the zinc blende[35] structure in epitaxial films, and with the rock salt structure at pressure above ~9 GPa.[34] ZnO has multiple potential applications in various technological fields, including optoelectronic, transparent electrode and spintronic devices.[34] However, defects cause a residual intrinsic *n*-type conductivity in ZnO, which affects the electrical and optical performance and hinders the numerous potential applications.[36] Thus, the use of ZnO in technological devices requires controlling defects during growth and device processing. However, the nature of the defect behind the persistent *n*-type conductivity remains controversial despite decades of research. Different donor-like defects have been proposed based on an outstanding number of DFT calculations,[37–39] e.g., hydrogen impurities,[40] a metastable oxygen vacancy,[41] complexes of zinc interstitial with nitrogen impurity[42] or oxygen vacancy,[43] silicon impurities,[44] complexes of hydrogen impurities with other intrinsic and extrinsic defects,[45] and carbon impurities.[46]

These DFT results are not sufficiently reliable due to the band-gap problem. Different approaches, including empirical and semi-empirical approximations, have been proposed to overcome the band-gap problem, e.g., extrapolation schemes, (LDA/GGA)+U, modification of pseudopotentials, fitting of hybrid functionals and DFT followed by quasiparticle calculations (DFT+GW).[33] However, the different theories can lead to scattered and contradictory results. In the case of the oxygen vacancy ($V_O$) in ZnO, where many of these approaches have been applied, the result is an ongoing controversy about the formation energy and the position of the thermodynamic transition levels in the band-gap.[39] For instance, the formation energy of the neutral $V_O$ ranges from 3.2 to 6.7 eV (under O-rich conditions) for calculations with different DFT approximations. Similarly, the transition level 2+/0 ranges from 0.3 to 2.8 eV.[47]

In this work, we present the application of DMC to study the structural stability and the energetics of defects in transition-metal oxides. As a benchmark, we have evaluated the equation of state of Zn crystal and ZnO in various phases and studied the energetics of the oxygen vacancy in ZnO. We also applied DMC to study the more challenging defects of zinc interstitials and hydrogen impurities in ZnO. The DMC calculations of the different phases agree with experiment to within 0.3 eV, indicating that the DMC calculations are reliable at the current level of approximation. Furthermore, extensive calculations show that the errors in our DMC results for oxygen vacancy are of the order of 0.2 eV. The calculations, therefore, establish a baseline value for the formation energy and the position of transition levels for $V_O$ in ZnO. Moreover, DMC indicates that the formation energies of hydrogen impurities and zinc interstitials in ZnO are above 1.4 eV under *n*-type and Zn- and H-rich conditions, indicating that their concentrations in ZnO under these conditions is low.

The rest of the paper is organized as follows. In Sec. II, we give a brief overview of the DMC method and computational details, including the validation of the pseudopotentials employed in our DMC calculations. In Sec. III, we show and discuss our main results. We start with a comparative study of DMC and a hybrid functional for the bulk properties of Zn and ZnO. Afterward, we discuss preliminary results to explore the expected level of accuracy for our DMC calculation of defects. We later discuss the energetics of the oxygen vacancy in ZnO in comparison with results from experiments and various DFT approximations. At the end of Sec. III, we discuss DMC results for hydrogen impurities and zinc interstitials in ZnO. We conclude in Sec. IV with a summary of our calculations and findings.

## II. METHODS AND VALIDATIONS

### A. DMC

DMC is a stochastic projector method that projects out the ground state solution of the many-body Schrödinger equation by iteratively applying an imaginary time Green's function to a trial wave function $\Psi_T$. It is one of the most accurate methods available for large quantum many-body problems. Compared to quantum-chemical many-body methods, DMC has a lower computational cost and more levels of parallelization[13] which makes it the only practical method for large complex materials. DMC methods have been previously reviewed,[9,10,25,48–51] and here we only describe the main approximations in the method and the computational details of the present calculations. DMC calculations reported here were performed with QMCPACK.[13]

DMC requires some approximations to be practical. However, DMC has yielded unprecedented highly accurate results[9,10,25,50,51] because it can be systematically improved. First, the imaginary time propagation is done within the short time approximation, which becomes exact as the simulation imaginary time step is taken to zero ($\tau \rightarrow 0$).[9] This approximation calls for a compromise between simulation time and computational cost, and it requires using small time steps and analyzing the associated errors. In practice, this error is easily tested and corrected for (see sections II and III in the Supplemental Material[52] for full details). A second approximation is the fixed-node[9] approximation or the generalized fixed-phase approximation,[53] which is the most important approximation in DMC. For an $N$-electron system, the wave function is antisymmetric. Thus, if the wave function is real, it must have positive and negative parts separated by a $(3N – 1)$-dimensional nodal hypersurface. The fixed-node approximation is introduced to sample only the positive distribution of the $N$-electron wave function, avoiding crossings of the nodal surface and resultant sign changes, i.e., the fermion sign problem. The fixed-node approximation then gives exact results if the trial nodal surface is exact.[9] Otherwise, the DMC energy is an upper bound to the ground-state energy, i.e., fixed-node DMC is a variational method. Several methods have been developed to improve the nodal surface and, in turn, the DMC accuracy (see Ref. 54 and references in there). To partially account for fixed node error, we have considered nodal surfaces arising from LDA, Hubbard-corrected and hybrid functionals.

A third approximation is the use of pseudopotentials (usually nonlocal), which is necessary because the computational cost of DMC is proportional to $Z^{5.5-6.5}$, where $Z$ is the atomic number.[9] The use of nonlocal pseudopotentials in DMC introduces a sign problem analogous to the fermion sign problem.[9] An additional approximation is then required to perform DMC with nonlocal pseudopotentials. The most widely used approach is the locality approximation.[55] The DMC energy calculated with this approximation is no longer an upper bound to the ground-state energy. The errors from this approximation are in most cases smaller than the fixed-node error.[9] Other methods[56,57] are also available to deal with nonlocal pseudopotential within DMC, but they have been tested and implemented to a lesser extent. The Zn pseudopotential used here has been carefully tested by comparing atomic, dimer, and bulk properties calculated with DMC to experimentally known values (see sections B and III.A for full details).

A final approximation, which does not directly come from the DMC algorithm, is the use of supercells to simulate condensed matter. This approximation introduces finite-size (FS) errors in all QMC calculations.[58] These errors are common to most many-body methods. These errors can be divided into one and two-body FS errors. One-body errors come from incorrect momentum quantization due to confinement of electrons in the simulation cell. The two-body error arises from the artificial periodicity of the exchange-correlation hole.[58] These FS errors decrease with the size of the supercell and are usually eliminated by extrapolation techniques. Various approaches are now available to reduce the need for extrapolation.[58–62] In the following sections, we describe the employed pseudopotentials and our DMC calculations of the equation of state (EOS) of Zn and ZnO crystals and the formation energy of defects in ZnO.

### B. Pseudopotentials

Ne- and He-core norm-conserving pseudopotentials (PPs) were used for the Zn and O atom, respectively. Our Zn-PP is based on the 20-electron core PP proposed in Ref. 63. We modified[64] the Zn-PP slightly by adding scalar-relativistic corrections.[65] We used the optimization method of Ref. 66 to converge the total energy to 8 meV while keeping a plane-wave energy cutoff of 300 Ry. The PPs were tested with DMC by evaluating the ionization potentials (IP) of the Zn and O atoms and the equilibrium distances ($r_e$) and dissociation energies ($D_e$) of the ZnO and $O_2$ dimers. An imaginary time step of 0.0025 $Ha^{-1}$ and 8192 walkers were sufficient to converge total energies to within 0.01 eV. Single-particle were generated with the plane-wave based code Quantum ESPRESSO[67] within LDA (unless otherwise specified) and orthorhombic cells with repeated images separated by more than 1.3 nm.

The results of these test calculations are included in **Table 1**. DMC yields IP and $D_e$ within 0.15 eV of the experimental values. CCSD(T) calculations[68–70] also produce similar deviations. The DMC equilibrium distances $r_e$ of the dimers agree well with experiments, to within 0.01 Å (data not shown). The small deviation of DMC for IP and $D_e$ can come from various sources, e.g., incomplete treatment of relativistic effects, fixed-node approximation, nonlocal PP approximation in DMC and missing core-valence electron correlation from the PP approximation.[71] We performed various calculations to analyze these errors for the IP of Zn. These calculations indicate that the errors for the IP of Zn calculated with DMC and PP come mainly from fixed-node errors and the PP approximation (see the Supplemental Material[52] for details). More accurate wavefunctions (e.g., multideterminant wavefunctions[72] and germinal products[73,74]) and alternatives to the use of pseudopotentials are required to reduce these errors and reach a higher accuracy with the fixed-phase DMC method.

**Table 1.** Ionization potential (IP) of Zn and O and dissociation energy ($D_e$) of ZnO and $O_2$ dimers calculated with DMC within the locality approximation.[a] Experimental results[75–77] and previous CCSD(T) calculations[68–70] are also included.

| Property | System | DMC | CCSD(T) | Exp. |
| --- | --- | --- | --- | --- |
| IP (eV) | O | 13.628(3) | 13.514 | 13.6181 |
|  | Zn | 9.257(9) | 9.402 | 9.3942 |
| $D_e$ (eV) | $O_2$ | 4.989(9) | 5.120 | 5.117 |
|  | ZnO | 1.28(1) | 1.63 | 1.44 |

[a]The DMC error is given in parentheses. The dissociation energies are relative to the ground state of the atomic species. As ZnO dissociates to $Zn(^1S) + O(^1D)$, instead of $Zn(^1S) + O(^3P)$, the experimental dissociation energy of ZnO (3.41 eV)[76] was shifted by the $O(^3P) - O(^1D)$ measured energy difference, 1.967 eV.[75]

### C.  Equation of state

The EOS of ZnO in the rock salt (B1), zinc blende (B3) and wurtzite (B4) phases, and Zn crystal were evaluated with DMC. The calculations were performed within the locality approximation.[9] For comparison, calculations were also performed with the scheme proposed by Casula,[56] the so-called T-moves approach. We used single-determinant Slater-Jastrow trial wavefunctions with one-, two- and tree-body Jastrow factors. Parameters in the Jastrow factors were optimized with variance minimization.[78] The single-particle orbitals populating the Slater determinant were generated with the plane-wave based code Quantum ESPRESSO[67] within LDA. The plane-wave energy cutoff was set to 4082 eV (300 Ry) because hard pseudopotentials were used. The number of walkers in the DMC simulations was 1024 or more, which was enough to reduce population errors to 0.01 eV per formula unit (f.u.).

Time step and FS errors were analyzed for each system. The details and results of these analyses are included as Supplemental Material.[52] We also explored fixed-node errors by generating single-particle orbitals with Hubbard-corrected and Heyd-Scuseria-Ernzerhof (HSE)[79] hybrid functionals. For the Zn crystal, the DMC energy evaluated with single-particle orbitals from LDA, HSE and LDA+$U_d$ with $U_d$ = 4, 8, 12 and 16 eV, were similar to with 0.03(1) eV/f.u. For ZnO (B4 phase), the DMC energy evaluated with single-particle orbitals from LDA and HSE were similar to within 0.063(6) eV/f.u. The DMC energies are lower with orbitals form

LDA+U$_d$ or HSE. The Zn crystal and wurtzite phase of ZnO have hexagonal structures. Ideally, the EOS of these structures will be calculated with a two-dimensional search.[80] However, we evaluated the EOS of hexagonal structures approximately, with a one-dimensional search, by fixing the *c/a* ratio to the experimental value.

### D.  Ionic defects

To study donor-like defects in the wurtzite phase of ZnO, we evaluated the formation energy of the oxygen vacancy ($V_O$) and the Zn interstitial at the octahedral site ($Zn_i$).[81] We also studied various forms of hydrogen impurity:[40] *i)* interstitial hydrogen centered at the Zn-O bond along the *c* axis ($H_i$), *ii)* substitutional hydrogen on an oxygen site ($H_O$), and *iii)* molecular hydrogen in the interstitial channel orientated along the *c* axis ($H_{2,int}$). The formation energy is defined as[33]

$$E^f[X^q] = E_{tot}[X^q] - E_{tot}[ZnO] - \sum_i n_i \mu_i + qE_t + E_{corr}[X^q] \quad (1),$$

where $X^q$ is a defect with charge $q$. $E_{tot}[X^q]$ and $E_{tot}[ZnO]$ are the total energies of ZnO with a defect and the equivalent bulk ZnO, respectively. $n_i$ indicates the number of atoms of type *i* that are added ($n_i > 0$) or removed ($n_i < 0$), and $\mu_i$ is the chemical potential of the atom *i*. $E_t$ is the Fermi energy relative to the valence-band maximum (VBM) of ZnO. $E_{corr}[X^q]$ is a correction term to account for supercell size effects, such as defect-level dispersion, electrostatic and elastic interactions.[33] The chemical potentials $\mu_O$ and $\mu_{Zn}$ were set to the O-poor (Zn-rich)

$[\mu_{Zn} = E_{tot}[Zn]$ and $\mu_O = \frac{1}{2}E_{tot}[O_2] + \Delta H_f]$ and the O-rich (Zn-poor) limits

$[\mu_{Zn} = E_{tot}[Zn] + \Delta H_f$ and $\mu_O = \frac{1}{2}E_{tot}[O_2]]$. $\Delta H_f$ is the formation enthalpy of ZnO, and $E_{tot}[Zn]$ and $E_{tot}[O_2]$ are the total energy of Zn crystal (per f.u.) and $O_2$, respectively. For hydrogen, $\mu_H = \frac{1}{2}E_{tot}[H_2]$ was taken as the H-rich limit.

We evaluated $E_{tot}[X^q]$ and $E_{tot}[ZnO]$ with DMC employing the T-moves approach. For $V_O$, DMC calculations were also performed within the locality approximation. Single-particle orbitals were generated within LDA. The total DMC energies of ZnO with and without a defect (e.g., $V_O^0$) are, respectively, 0.059(6) and 0.063(6) eV/f.u. lower if orbitals are generated with HSE instead of LDA. Fixed-node errors mostly cancel out for $E^f[V_O^0]$. We used an imaginary time step of 0.02 Ha$^{-1}$ for defect calculations; this time step results in a converged $E^f[V_O^0]$ due to time step error cancelation (see the Supplemental Material[52] for details). Defected and bulk ZnO was simulated with a 32 atom supercell with the experimental lattice constants ($a$ = 3.242 Å, $c$ = 5.188 Å)[82] of the wurtzite phase and twisted boundary conditions on a 4×4×2 grid. This grid was sufficient to converge $E^f[V_O^0]$ to within 0.1 eV. Test calculations were also performed for $V_O$ with a larger 48 atom supercell on a 2×2×2 grid to explore finite size effects.

The atomic positions of the defects were optimized within LDA until residual forces were below 0.02 eV/Å. Today, it is impractical to calculate interatomic forces in large systems with DMC, and atomic configurations are usually taken from a DFT approximation. However, for energy differences (e.g., defect formation energy), this approximation is expected to introduce only second order effects.[48] We have corroborated this expectation by evaluating the relaxation energy of $V_O^0$ with DMC and atomic position optimized with LDA and HSE. We also calculated the relaxation energy with various DFT approximations, i.e., LDA, PBE, PBE+U[83] and HSE. These DFT calculations were performed with the Vienna Ab-initio Software Package (VASP)[84–86] with projector augmented-wave (PAW)[87,88] potentials and an energy cutoff of 550 eV. DFT and DMC calculations were performed with a 32 atom supercell of wurtzite ZnO with experimental lattice constants on a 4×4×2 k-point (twist) grid. With experimental lattice constants, the internal structural parameter $u$ evaluated with the various DFT approximations was similar, i.e., 0.381; the experimental value[34] is 0.382.

We approximated the VBM and conduction band minimum (CBM) as the ionization potential (IP) ($[E(N) - E(N-1)]$) and the electron affinity (EA) ($[E(N+1) - E(N)]$), respectively, with respect to the average electrostatic potential within ZnO. The IP and EA were calculated at the Gamma-point to avoid significant finite size effects (e.g., band-filling effects)[33] and with various supercell sizes to study any residual size effects.

The correction term $E_{corr}[X^q]$ in Eq. (1) was evaluated within LDA. This term includes various types of corrections: i.e., dispersion of the defect level due to the overlap of the defect wavefunction with its neighboring image and electrostatic and elastic interactions. Defect-level dispersion errors for localized defect were corrected within first-order perturbation theory.[33] For delocalized (shallow) defects, dispersion errors were corrected using the dispersion of valence- or conduction-band.[33] For electrostatic and elastic interactions, we follow the correction model of Refs. 89 and 90. In this model, the interactions of the charged defect with its periodic image and background charge lead to $L^{-1}$, $L^{-3}$ and $L^{-5}$ dependencies of the formation energy as

$$E^f = E_0^f + \frac{\alpha q^2}{\varepsilon L} + \frac{\beta}{L^3} + O(L^{-5}) \qquad (2),$$

where $L$ is a representative supercell dimension. The $L^{-1}$ term is the Madelung energy of an array of point charge in a neutralizing background scaled by the macroscopic dielectric constant, $\varepsilon$. The formation energy at the diluted limit, $E_0^f$, is estimated from systematic calculations with supercells of increasing size. These calculations were performed with VASP and PAW potentials. We used Gamma-centered k-point meshes of 4×4×2 for 32 and 48 atom supercells and 2×2×2 for 72, 128, 192, and 256 atom supercells; these k-point meshes were enough to converge the total energy to less than 5 meV. The dielectric constant $\varepsilon$ was set to 8.1, which is the average of the measured[91] dielectric constant of ZnO. Our methodologies to evaluate $E_{corr}[X^q]$ closely follow the methods used in Ref. 92 and further details can be found there. Note that $E_{corr}^f[X^q]$ should also include correction for the two-body FS errors present in many-body theory within the supercell approximation.[58] For $E^f[V_O^q]$ evaluated with the 32 atom supercell, we estimated[58,62] these errors to be within the statistical error of the DMC calculations.

## III. RESULTS AND DISCUSSION

### A. Accurate bulk properties: a comparative study of DMC and screening adjusted HSE

The EOS of Zn and ZnO evaluated with DMC are shown in **Figure 1**. The derived structural parameters are included in **Table 2** and **Table 3** (the error bar of the parameters was estimated from the statistical uncertainties of the Monte Carlo data). Results from HSE38 (HSE with a fraction of 0.38 of nonlocal Fock-exchange)[92] are also included for comparison.

In general, the structural parameters evaluated with DMC and HSE38 for the Zn crystal are consistent with the experimental values (within 0.5%). Both methods also yield accurate values for the bulk modulus. For the first derivative of the bulk modulus, DMC yields a value in better agreement with experiment than HSE38. For the cohesive energy of Zn, DMC within the locality approximation has an error of ~0.35 eV. This error is reduced to ~0.1 eV with the T-moves approach. HSE38 has a similar error of ~0.1 eV for the cohesive energy of Zn. The error of ~0.35 eV contrasts with previous DMC calculations (within the locality approximation) of light metals (e.g. Li, Na, Mg, and Al),[10,23,93] where an accuracy of 0.1 eV was reached.

**Table 2.** Lattice constants *a* and *c* (Å), bulk modulus B (GPa) and its first derivative B′, and cohesive energy (eV) of Zn crystal evaluated with DMC and HSE38 (HSE with a fraction of 0.38 of nonlocal Fock-exchange). Experimental results[94,95] are included for comparison. The statistical uncertainty in DMC is provided in parenthesis.

| Method | *a, c* | B | B′ | Energy |
|---|---|---|---|---|
| DMC | 2.656(1), 4.931(1) | 70.2(4) | 5.22(4) | 1.01(2), 1.28(1)[a] |
| HSE38 | 2.677, 4.969 | 67.24 | 4.59 | 1.23 |
| Expt. | 2.6644, 4.9487 | 60 – 70 | 5.6 – 6.0 | 1.347 |

[a]DMC cohesive energy evaluated with the T-moves approach. The calculation was performed with the experimental equilibrium structure.

To the best of our knowledge, DMC has been previously applied to study only one other transition metal, Fe.[96] Those DMC calculations were performed within the locality approximation. However, the cohesive energy of Fe was not reported. In that work, the equation of state of Fe was calculated in the high-pressure region of the phase diagram, and the results agreed with experimental data.[96] Many-body finite-size effects for Fe were also studied, and the results showed that the method of Ref. 62 is very efficient to correct for two-body FS errors.[96] Our similar analysis of finite-size effects for Zn (see sections II in the Supplemental Material[52] for full details) agree with these findings.

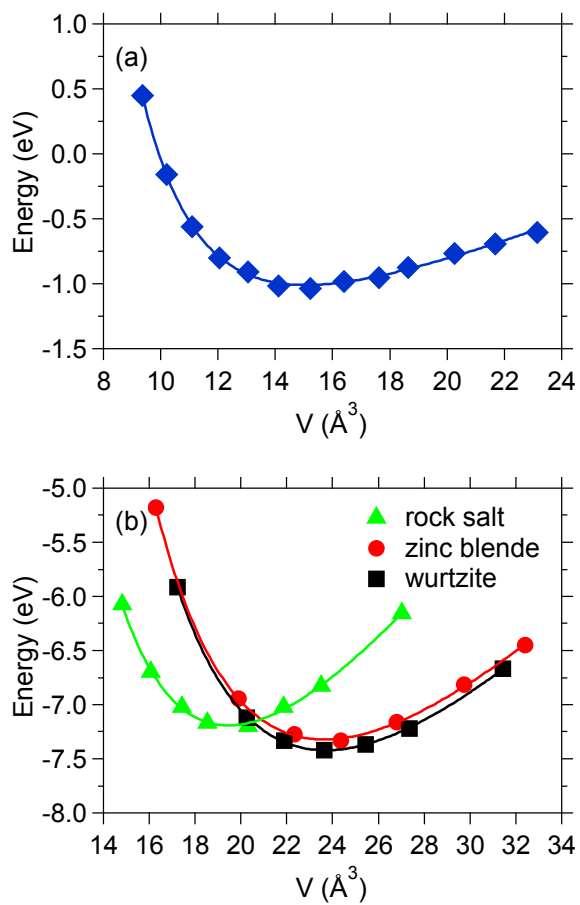

**Figure 1.** DMC energy versus volume per formula unit together with fitted Murnaghan curves for (a) Zn and (b) ZnO in the rock salt, zinc blende and wurtzite phases. Energies are per formula unit and relative to the Zn and O atoms. The statistical uncertainty in DMC is smaller than the symbol size.

For ZnO, DMC and HSE38 yield the lattice constants of the B1 and B4 phases within 0.4% of the experimental value (**Table 3**). In the case of the B3 phase, DMC and HSE38 overestimate the lattice constant by more than 3%. The B3 phase of ZnO can be stabilized only by growing on cubic substrates.[35] The significant deviation of the calculated lattice constant comes from the effect of strain in the experimental heterostructure.[35] For the B1 and B4 phases, the calculated bulk moduli are in agreement with experimental measurements: no experimental data is available for the B3 phase. For the pressure derivate of the bulk modulus, both methods slightly overestimate the experimental values.

**Table 3.** Lattice constants *a* and *c* (Å), bulk moduli B (GPa) and their first derivative B′, and the cohesive energies (eV) of ZnO in the rock salt (B1), zinc blende (B3), and wurtzite (B4) phases evaluated with DMC and HSE38. Experimental results are included for comparison (see Refs. 34,35,82,97). The statistical uncertainty in DMC is provided in parenthesis.

| Method | Phase | Lattice *a*, *c* | B | B′ | Energy |
|---|---|---|---|---|---|
| DMC | B1 | 4.268(1) | 194.2(5) | 4.28(5) | 7.19(2) |
| HSE38 | | 4.268 | 190.3 | 4.38 | 6.65 |
| Expt. | | 4.271 – 4.294 | 202.5, 228 | 3.54, 4.0 | |
| DMC | B3 | 4.556(1) | 147.2(4) | 4.32(3) | 7.32(2), 7.67(2)[a] |
| HSE38 | | 4.571 | 143.3 | 4.12 | 6.84 |
| Expt. | | 4.18, 4.37 – 4.47 | | | |
| DMC | B4 | 3.245(1), 5.193(1) | 151.6(4) | 4.21(4) | 7.42(2), 7.80(2)[a] |
| HSE38 | | 3.254, 5.208 | 144.6 | 4.15 | 6.87 |
| Expt. | | 3.242, 5.188 | 140 – 170 | 3.6, 4.0 | 7.52 |

[a]DMC cohesive energy evaluated with the T-moves approach and corrected for 2-body FS errors[62] and residual FS errors. Calculations were performed with the experimental equilibrium structures.

The cohesive energies evaluated with DMC and HSE38 for ZnO in the B1, B3 and B4 phases are included in **Table 3**. Both methods correctly predict the stability order of the various phases, B4 > B3 > B1. The energy separation between the B4 and B1 phases is essentially the same in both methods, 0.2 eV. On the other hand, the energy difference between the B4 and B3 phases is ~0.1 eV larger in DMC than HSE38. The experimental cohesive energy is only available for the B4 (7.52 eV).[34] The cohesive energy of this phase evaluated with DMC within the locality approximation and the T-moves approach is 7.42(2) and 7.80(2) eV, respectively (**Table 3**). Both of these results are in reasonable agreement with the measured value, deviating only by -0.10(2) and 0.28(2) eV. In contrast, HSE38 underestimates the cohesive energy of B4 by 0.65 eV.

The structural transition of ZnO from the B4 to the B1 phase have been observed at pressures between 8.7 and 10 GPa with a volume change of -3.98 – -4.31 Å$^3$.[34] The value of the transition pressure $p_{T1}(B4 \rightarrow B1)$ at finite temperature can be evaluated by calculating the pressure where the Gibbs free energy is equal for the two phases. At $T = 0$, the Gibbs free energy reduces to the enthalpy $H(p) = E_0 + pV(p)$. We determined $V(p)$ by inverting the Murnaghan expression.[98] $p_{T1}(B4 \rightarrow B1)$ evaluated with DMC, HSE38 and LDA at $T = 0$ is 9.17(8), 8.4, and 8.9 GPa, respectively. The corresponding calculated volume change $\Delta V(B4 \rightarrow B1)$ is -3.82(2), -4.00 and -3.63 Å$^3$. The DMC and HSE38 results are close to the experimental values. On the other hand, LDA yields a $p_{T1}(B4 \rightarrow B1)$ value that falls within the experimental range, but it underestimates $\Delta V(B4 \rightarrow B1)$.

From bulk Zn, ZnO and $O_2$, we can estimate the enthalpy of formation of ZnO. The value evaluated with DMC within the locality approximation and the T-moves approach for the B4 phase is -3.94(2) and -4.15(2) eV, respectively. The corresponding value evaluated with HSE38 is -3.25 eV, which is similar to previous HSE calculations.[92,99] The experimental enthalpy of formation of the B4 phase is -3.63 eV.[34] DMC and HSE38 have large error for the enthalpy of formation. The large errors are not unexpected because calculation of the enthalpy of formation of ZnO requires total energy differences between an insulator, a pure metal, and molecular $O_2$. For accurate enthalpy of formation, total energies for all these phases need to be highly accurate to avoid error accumulation.

## B. Preliminaries for defects: expected level of accuracy

These preliminary calculations aim to explore the expected level of accuracy for our DMC calculations of defects. We first discuss DMC results for the quasi-particle energy gap and the relaxation energy of $V_O^q$ in ZnO. Accurate quasi-particle energy gap and atomic positions are important for charged defects. We also discuss the expected fixed-node errors and errors from fine size effects for defects (the $E_{corr}[X^q]$ term in Eq. (1)).

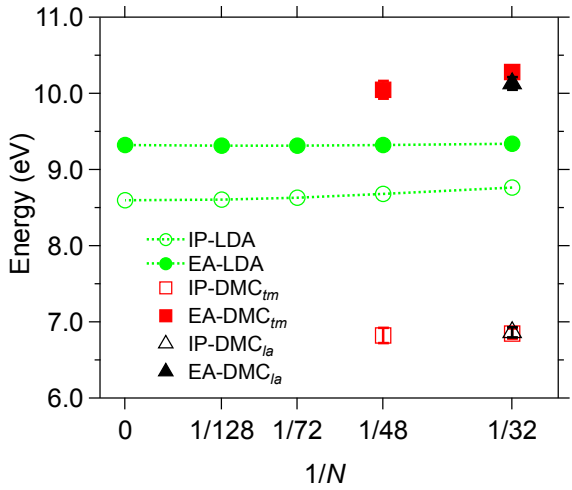

**Figure 2**. Ionization potential (IP) and electron affinity (EA) of wurtzite ZnO calculated with LDA at the Gamma-point for various supercell sizes. The VBM and CBM from LDA are included at $1/N = 0$. The IP and EA evaluated with DMC within the locality approximation ($DMC_{la}$) and the T-moves approach ($DMC_{tm}$) are also shown. Note the overlapping symbols for the values at $1/N = 1/32$. Standard deviation error bars show statistical uncertainties in the DMC data.

**Figure 2** displays the IP and EA of wurtzite ZnO evaluated with LDA and DMC as a function of the inverse supercell size. Residual size effects are small within LDA. The IP evaluated with LDA is slightly high for the smaller supercell while the EA is similar for all supercell sizes. In DMC, the IP is similar for the two supercells that were explored. These results indicate that IP ≈ VBM is a good approximation for ZnO simulated with 32 and 48 atom supercells. The EA is slightly different for the 32 and 48 atom supercells in DMC. Similarly, the EA is somewhat lower for DMC within the locality approximation than the T-moves approach. The differences, however, are small and within a statistical error of 0.15 eV. From the DMC IP and EA, the quasi-particle gap for ZnO is estimated at 3.26(10) eV within the locality approximation and the 32 atom supercell. DMC with the T-moves approach and 32 and 48 atom supercells yield 3.43(9) and 3.22(16) eV, respectively, which are similar within a statistical error of 0.18 eV. These DMC values for the quasi-particle gap of ZnO are in close agreement with the experimental band-gap of 3.44 eV.[100] Our results also agree with the band gap of ZnO recently evaluated with the auxiliary-field QMC method,[28] i.e. 3.26(16). Close agreement between DMC quasi-particle gap and experimental band gap have also been reported for MgO,[22] $TiO_2$,[18] FeO,[101] MnO,[102] and $La_2CuO_4$ and $CaCuO_2$.[16]

**Figure 3** displays the relaxation energy of $V_O^q$ evaluated with various DFT approximations and DMC. As previously shown by Ágoston and Albe,[103] the relaxation energies of these defects are sensitive to the DFT approximations. The difference comes mainly from errors in the electronic properties. As shown in **Figure 3**, the relaxation energies of $V_O^q$ are similar whether atomic positions are optimized with LDA or each individual functional. The largest differences (~0.05 eV) are found for PBE, where the structure of the vacancy optimized with LDA and PBE differs; for instance, by ~0.1 Å for $V_O^{2+}$. The defect structure optimized with LDA, PBE+U and HSE are similar. The maximum differences between the atomic positions of $V_O^{2+}$ in ZnO optimized with HSE38 (HSE method with a 0.375 fraction of nonlocal Fock-exchange)[92] and LDA, are of the order of 0.02-0.04 Å, while the average difference is only 0.01 Å. Therefore, the error introduced in our DMC calculations by using atomic positions optimized with LDA is small. In fact, the DMC formation energy of $V_O^{2+}$ evaluated with atomic positions optimized HSE38 and LDA differ only by 0.01(17) eV.

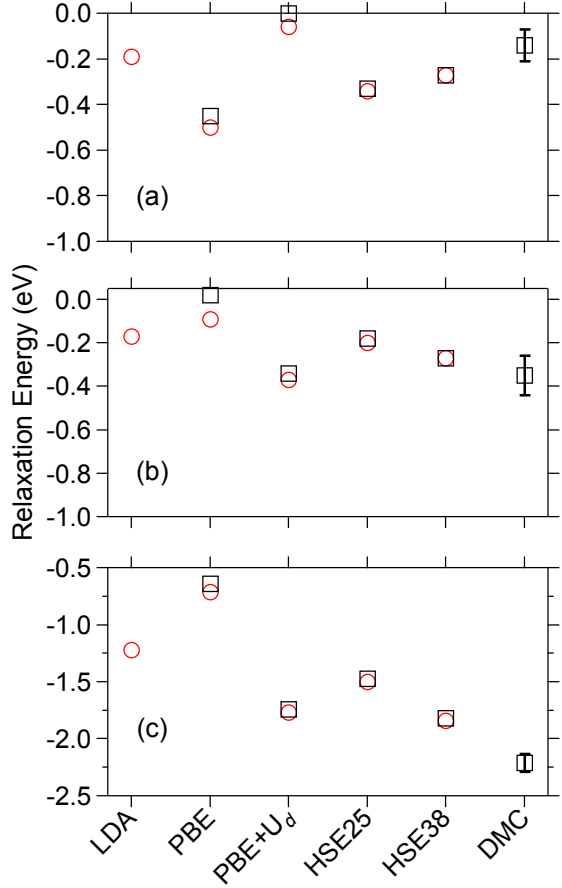

**Figure 3.** Relaxation energy of (a) $V_O^0$, (b) $V_O^{1+}$ and (c) $V_O^{2+}$ in ZnO evaluated with various DFT approximations and DMC. The circle symbols correspond to calculations with atomic positions optimized with each DFT approximation. The square symbols correspond to calculations with atomic positions optimized with LDA. HSE25 and HSE38 correspond to the HSE method with a 0.25 and 0.375 fraction of nonlocal Fock-exchange, respectively. Standard deviation error bars show statistical uncertainties in the DMC data.

Based on the calculated cohesive energy of Zn and ZnO, the overall accuracy of our DMC calculations is ~0.3 eV. These errors, which are likely fixed-node errors, mostly cancel out for energy differences between similar systems. For instance, fixed-node errors are very similar in bulk and defected ZnO (e.g., with $V_O^0$) as mentioned before. Therefore, defect formation energies are not expected to have the significant DMC errors found for the enthalpy of formation of ZnO.

This is expected particularly for $V_O^q$ and the hydrogen impurities because fixed-node errors are relatively small for the chemical potentials of oxygen and hydrogen. For $Zn_i$, the defect formation energy will directly reflect any DMC errors in bulk Zn since the DMC energy sets the chemical potential of Zn. However, these errors are ~0.1 eV when using the T-moves approach, which is the method that we employed to evaluate defect formation energies with DMC.

Using LDA to estimate $E_{corr}[X^q]$ introduces an uncertainty in our DMC calculations of defects because the degree to which LDA delocalizes electronic states is different from DMC. This is a necessary approximation because it is impractical at the moment to perform DMC calculations for transition metal oxides with supercells large enough to estimate these errors. Similar approaches have been employed previously to study the oxygen vacancy in MgO with DMC[22] and various ionic defects in ZnO with hybrid functionals.[92,99] We do not expect any qualitative difference due to this approximation for $V_O$. The oxygen vacancy in ZnO is a localized defect and size effects are expected to be small for $E^f(V_O^0)$.[104] Size effects are more relevant in the charged vacancy because of spurious electrostatic interactions. As shown in the next section, we calculated $E^f(V_O^q)$ with DMC employing 32 and 48 atom supercells and applying $E_{corr}[V_O^q]$ estimated with LDA. The two set of calculations yield formation energies that are similar within statistical errors below 0.2 eV. The situation will likely be different for delocalized defects (e.g., $Zn_i^{0,1+}$, $H_i^0$, $H_O^0$). In these cases, $E_{corr}[X^q]$ is evaluated from the dispersion of the valence- or conduction-band, which could certainly be different[28] in LDA and DMC. However, this large uncertainty is not present for $E_{corr}[X^q]$ of fully ionized states ($Zn_i^{2+}$, $H_i^{1+}$, $H_O^{1+}$). The current model can bring valuable insight into the energetic of these defects.

### C. Energetics of $V_O$ in ZnO from DMC

We include in **Table 4** DMC results for the formation energy, thermodynamic transition levels and U effective (see below) for the oxygen vacancy under O-rich conditions and $E_F$ = VBM. The DMC formation energy of the neutral state ($V_O^0$) is 4.81(7) and 5.02(8) eV when evaluated in the 32 atom supercell within the locality approximation and the T-moves approach, respectively. Calculations with the T-moves approach and a 48 atom supercell yield a value similar within an statistical error of 0.14 eV. Note that for $V_O^{1+}$ and $V_O^{2+}$, DMC calculations with the 32 and 48 atom supercells also yield statistically similar formation energies. The formation energies evaluated with the locality approximation and the T-moves approach are similar, in part due to error cancelation. However, we focus our discussion on the results from the T-moves approach. The available experimental data on $E^f(V_O)$ is limited because such measurements are indirect and depend on defect identification. We found two estimates in the literature for $E^f(V_O)$ in ZnO. The first estimate is ~6 eV.[105] A second estimate can be found in Ref. 106, i.e. 1.9 eV under Zn-rich conditions or 5.5 eV if converted to the O-rich limit with the experimental formation enthalpy of ZnO. The authors in Ref. 106 did not associate the measured value directly with $V_O$ as it was much higher than their DFT calculations.

Using $E^f(V_O)$, the concentration $[V_O]$ in ZnO can be estimated and compared with available experimental results. $[V_O]$ in as-grown ZnO at temperature of 1373 K and under Zn-rich conditions was estimated to be close to $10^{17}$ cm$^{-3}$ from positron annihilation spectroscopy measurements.[107,108] Taking Zn-rich conditions by converting the DMC results for $E^f(V_O)$ with the DMC formation enthalpy and using a thermal Boltzmann distribution, we obtain $[V_O]$ = 1 - 6 x $10^{19}$ cm$^{-3}$ at 1373 K. This concentration deviates from the experimental estimate because our DMC calculations overestimated the formation enthalpy of ZnO due to error accumulation (see discussion above). If we use the experimental formation enthalpy of ZnO instead of the DMC value to transform to Zn-rich conditions, $[V_O]$ is 1 - 40 x $10^{17}$ cm$^{-3}$. (The ranges of $[V_O]$ reflect the statistical errors in the DMC results, as well as the difference between the locality approximation and the T-moves approach for $E^f(V_O)$). For comparison, the corresponding $[V_O]$ evaluated with $E^f(V_O)$ from HSE38 under Zn-rich conditions is 3 x $10^{19}$ cm$^{-3}$. As we mentioned before, HSE38 also has a significant error for the formation enthalpy of ZnO. Using the experimental formation enthalpy of ZnO instead of the HSE38 value to transform to Zn-rich conditions yield $[V_O]$ = 7 x $10^{20}$ cm$^{-3}$. $[V_O]$ evaluated with HSE38 is much larger than the experimental estimate, indicating that $E^f(V_O)$ calculated with HSE38 is too low (see next section for further comparisons of DMC and HSE).

**Table 4.** Oxygen vacancy formation energy $E^f(V_O^q)$, thermodynamic transition levels $\varepsilon(q/q')$, and U effective $U_{eff}$ in ZnO. DMC calculations were performed with the locality approximation (LA) and the T-moves approach (TM) in 32 and 48 atom supercells under O-rich conditions and $E_F$ = VBM, all values in eV. Finite-size corrections due to image interactions were approximated within LDA.[a] The DMC statistical error is included in parenthesis.

| Method | Supercell | $E^f(V_O^0)$ | $E^f(V_O^{1+})$ | $E^f(V_O^{2+})$ | $\varepsilon(2+/1+)$ | $\varepsilon(2+/0)$ | $\varepsilon(1+/0)$ | $U_{eff}$ |
|---|---|---|---|---|---|---|---|---|
| LA | 32 | 4.81(7) | 2.70(9) | 0.75(11) | 1.95(9) | 2.03(11) | 2.11(9) | 0.17(10) |
| TM | 32 | 5.02(8) | 2.77(11) | 0.97(13) | 1.80(10) | 2.03(12) | 2.25(10) | 0.45(10) |
|  | 48 | 5.17(11) | 2.69(13) | 0.82(17) | 1.87(14) | 2.18(18) | 2.48(15) | 0.61(14) |

[a]The correction terms $E_{corr}(V_O^q)$ applied to the DMC formation energy of $V_O^0$, $V_O^{1+}$ and $V_O^{2+}$ evaluated with the 32 supercell are -0.20, -0.02 and 0.11 eV, respectively. The corresponding values for calculations with the 48 atom supercell are -0.13, 0.07 and 0.40 eV.

DMC predicts the $V_O$ as a deep donor, with the transition levels (2+/1+), (2+/0) and (1+/0) located at 1.8 – 2.5 eV above the VBM (**Table 4**). Transition levels of $V_O$ in ZnO have been studied with various experimental techniques.[109–113] The measured transition levels that have been associated with $V_O$ are scattered and can be grouped according to their position from the VBM: (*i*) 1 - 1.3 eV, (*ii*) 2.2 - 2.5 eV and (*iii*) 2.9 - 3.29 eV. Our DMC results indicate that the transition levels in groups (*i*) and (*iii*) do not correspond to the isolated $V_O$ in ZnO.

The disproportion between the singly, neutral and doubly charged $V_O$ can be quantified by the effective U parameter ($U_{eff}$).[33] $U_{eff}$ quantifies the repulsive electrostatic interaction between electrons in defect states. It has electronic ($U_{el}$) and atomic relaxation ($U_{rel}$) contributions, i.e. $U_{eff} \approx U_{el} + U_{rel}$. $U_{eff}$ and $U_{el}$ can be evaluated as:

$$U = E_{tot}^d(V_O^{2-}) + E_{tot}^d(V_O^0) - 2E_{tot}^d(V_O^{1+}) \qquad (3),$$

where $E_{tot}^d(V_O^q)$ corresponds to the total energy of $V_O^q$ at the dilute limit, with relaxed and fixed atomic structures for $U_{eff}$ and $U_{rel}$, respectively. To evaluate $U_{el}$, the atomic structure of $V_O^q$ was fixed to that of bulk ZnO. The $U_{eff}$ parameter is not the U value in DFT+U. $U_{eff}$ has a physical meaning[33] and can be experimentally observed. $U_{eff}$ evaluated with DMC is included in **Table 4**. DMC predicts a positive $U_{eff}$ value, above 0.1 eV. Note that calculations with the 48 atom cell yield a higher positive value. Based on experimental results,[114] it was recently suggested that $U_{eff}$ for $V_O$ in ZnO is negative. However, the measured transition levels at 2.9 and 3.29 eV in Ref. 114 that support the suggested negative U character of $V_O$ in ZnO are outside the range predicted by DMC. Moreover, the assignment of the transition at 2.9 eV to the (2+/0) level of an isolated $V_O$ is still controversial (see Ref. 110 for a detailed discussion).

D. **Analysis of quantitative differences between DMC and DFT approximations for $V_O$**

$V_O$ in ZnO has been widely studied with various DFT approximations.[37–39] To enable a consistent comparison with our DMC results, we also performed DFT calculations for $V_O$ and the band-gap in ZnO with a 32 atom supercell using LDA, GGA and hybrid functionals. In **Figure 4**, we compare our DMC and DFT calculations as well as previous DFT results. As expected, LDA and GGA approximations severely underestimate the band-gap of ZnO. (LDA/GGA)+U and hybrid functionals (e.g., HSE38) methods reproduce the band-gap of ZnO. However, these methods have been empirically parameterized. Only screened exchange functional (sX)[99] and GW can reproduce the band-gap without empirical parameters. For $E^f(V_O^q)$, the spread of the DFT results is large, coming mainly from the intrinsic limitations of the DFT approximations, and to a lesser extent from the different approximations employed to model dilute defect concentrations. None of the DFT calculations agree consistently with DMC. Deviations from DMC are over 1 eV for many DFT results, particularly for methods where the error in the band-gap is corrected with empirical and semi-empirical methods. $E^f(V_O^q)$ evaluated with hybrid functionals are the most consistent with each other. However, $E^f(V_O^0)$ evaluated with hybrid functionals deviates by over 0.5 eV from DMC within the locality approximation or the T-moves approach. As shown above, the DMC values are in better agreement with experimental results than hybrid functionals such as HSE38.

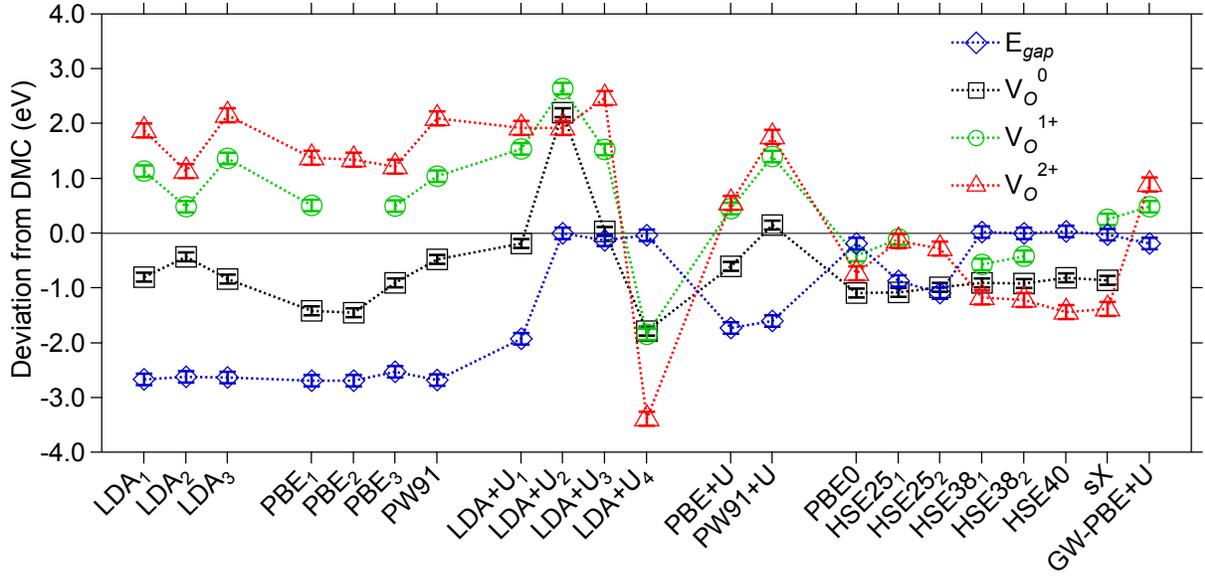

**Figure 4**. Deviation of DFT from DMC for the formation energy of the oxygen vacancy in ZnO under O-rich and $E_F = VBM$ conditions. Deviation of the band-gap is also included. References for DFT calculations are [$LDA_1$, $PBE_1$, PBE+U, PBE0, $HSE25_1$, $HSE38_1$] (present work), [$LDA_2$, $LDA+U_3$] (Ref. 115), [$LDA_3$, $LDA+U_1$, $LDA+U_2$] (Ref. 81), [$PBE_2$, $HSE38_2$] (Ref. 92), $PBE_3$ (Ref. 116), [PW91, PW91+U] (Ref. 83), $LDA+U_4$ (Ref. 47), [$HSE25_2$, HSE40] (Ref. 99), GW-PBE+U (Ref. 117). The DMC calculations were performed with the T-moves approach and a 32 atom supercell. Finite-size corrections due to image interactions in our DMC and DFT calculations were approximated within LDA (see Supplemental Material for details). Standard deviation error bars show statistical uncertainties in the DMC data.

The thermodynamic transition levels from DMC and hybrid functionals (data not shown) are in the same range, from 1.2 to 2.5 eV above the VBM. On the other hand, the parameter $U_{eff}$ is different in the two methods. DMC predicts $U_{eff}$ above 0.1 eV while for hybrid functionals and many other DFT approximations it is below -0.4 eV. Some previous calculations[83,117] have also predicted a positive or small negative value for $U_{eff}$. Note that $U_{eff}$ is particularly challenging to evaluate as it involves the equilibrium of electronic and relaxation effects, and within the supercell approximation it also requires proper description of finite size effects. However, the different value of $U_{eff}$ in DMC and DFT is associated with different descriptions of the electronic interactions in the two methods as quantified by $U_{el}$. $U_{el}$ evaluated with DMC and the 32 atom cell is 2.42(12) eV. For the various DFT approximations, $U_{el}$ ranges from 0.6 to 1.4 eV. On the other hand, $U_{rel}$ is rather similar in DMC and DFT. For instance, DMC and HSE38 yield -1.97(16) and -1.86 eV for $U_{rel}$, respectively.

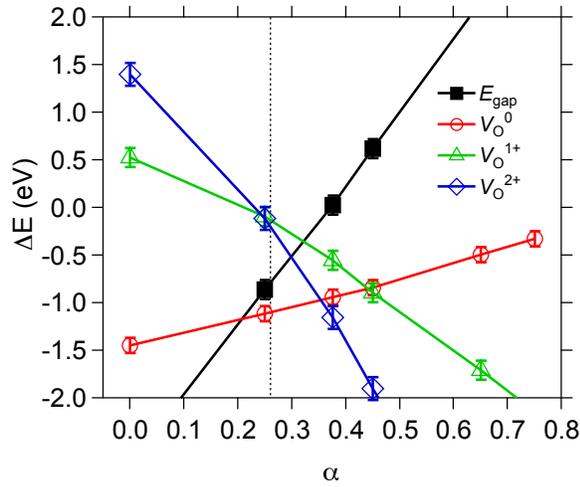

**Figure 5.** Deviation of HSE from DMC as a function of the fraction of nonlocal Fock-exchange ($\alpha$) in HSE for the band-gap energy ($E_{gap}$) and the formation energy of the oxygen vacancy ($V_O^q$) in ZnO. Formation energies evaluated at $E_F = VBM$ and under O-rich conditions. Results with PBE functional are included at $\alpha = 0$. The vertical dotted line indicates the $\alpha$ value (0.26) with the lower overall difference between DMC and HSE.

As shown above, defect formation energy evaluated with DMC and DFT approximations can differ significantly due to the intrinsic limitations of these approximations to describe electronic interactions in transition-metal oxides. A common approach to improve the incorrect description of electronic interactions in DFT is to fit the fraction of nonlocal Fock-exchange in hybrid functionals to reproduce measured properties, e.g., band-gap energy.[33] **Figure 5** shows the deviation of HSE from DMC for the band-gap energy and $E^f(V_O^q)$ in ZnO as a function the fraction of nonlocal Fock-exchange ($\alpha$) in HSE. There is not a unique $\alpha$ value that reproduces all DMC results. The value of $\alpha$ with the lowest error for the band-gap and $E^f(V_O^q)$ is $\alpha = 0.26$.

DMC has been previously applied to study ionic defects in various materials and the results compared with local and semi-local DFT approximations.[22–24] For instance, formation energies evaluated with DMC are higher (by 0.3 - 1 eV) than GGA for neutral oxygen vacancy in MgO[22] and self-interstitial in aluminum[23] and silicon.[24] The main error of GGA for $V_O^q$ in MgO was attributed to the overly delocalized description in GGA, which results in a small and underestimated band gap for MgO.[22] As a result of the small, the deep doubly occupied $V_O^q$ level is located too close to the VBM, resulting in low formation energy.[22] A similar argument can explain the fact that LDA and GGA yield lower formation energies than DMC for $V_O^q$ in ZnO (**Figure 4**). Correcting the band gap with a hybrid functional like HSE should then improve the agreement with DMC. Indeed, using HSE with $\alpha = 38$ yields $E^f(V_O^0)$ 0.5 eV closer to DMC than PBE-GGA (**Figure 5**). However, the difference with DMC is still well over 0.5 eV for any sensible fraction of nonlocal Fock-exchange in HSE (e.g., $0.3 < \alpha < 0.45$). The results shown above for $U_{eff}$ suggest that this error is due to the different descriptions of the electronic interactions in the DMC and DFT methods. However, further analysis is required for a better understanding of the source of the DFT error in the case of $V_O^q$ in ZnO.

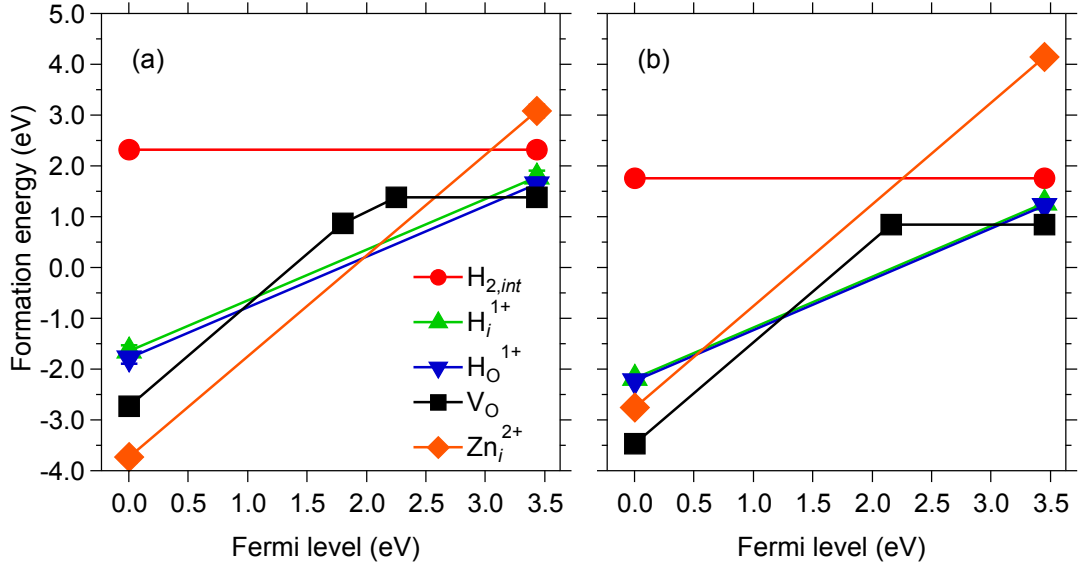

**Figure 6**. Formation energy of ionic defects in ZnO evaluated with (a) DMC with the T-moves approach and (b) HSE38 as a function of the Fermi energy under Zn-rich conditions. Under these conditions, the DMC formation energy of $V_O$ and $H_O$ were calculated using the experimental enthalpy of formation of ZnO. The statistical uncertainty in DMC is smaller than the symbol size.

### E. Shallow donors in ZnO from DMC

**Figure 6** shows formation energies of various donor-like defects in ZnO evaluated with DMC and HSE38, with correction for finite-size effects, as a function of the Fermi energy under Zn-rich conditions. These calculations were performed with a 32 atom supercell and $E_{corr}[X^q]$ estimated with LDA. Note that our HSE38 results with this model resemble very closely previous HSE38 calculations with a larger 192 atom supercell.[92] Only charged states that are stable in the given range of Fermi energy are included in **Figure 6**. The Fermi energies where the slopes change indicate the thermodynamic transition levels. As discussed above, the oxygen vacancy is a deep donor with transition levels located below 2.5 eV from the VBM. For the Zn interstitial $Zn_i$, hydrogen interstitial $H_i$ and $H_O$, DMC predicts the transition levels to be above the CBM (**Table 5**). Different DFT studies,[37–39] as well as experimental results,[36] indicate that these defects are shallow donor with transition levels slightly below the CBM. DMC yields the levels above the CBM likely because of uncorrected image interaction errors. As mentioned previously, these interactions come from dispersion of the defect-level due to the overlap of the defect wavefunction with its neighboring image.[33] The correction term $E'_{corr}[X^q]$, which is estimated from the dispersion of the CBM within LDA for shallow donors, fails to remove all size effects for these defects in the DMC calculations. Shallow defects require supercell larger than the present 32 atom model. Alternatively, errors from defect-level overlapping in shallow defects could be estimated more accurately by evaluating the band structure directly with DMC. Albeit computationally demanding for a material like ZnO, calculations of band structure with QMC methods are practical.[28]

The errors due to defect-level overlapping in shallow defects are present in the neutral and partially ionized states. For $Zn_i$, $H_i$ and $H_o$ in ZnO, we can use the formation energies of the fully ionized states to estimate the energetics of these defects under *n*-type conditions (Fermi energy near the CBM). The DMC formation energies of the fully ionized charged states have the uncertainty introduced by estimating $E'_{corr}[X^q]$ with LDA. However, our test calculations with 32 and 48 atom supercells for the charged states of the oxygen vacancy indicate that this a reasonable approach to estimate $E'_{corr}[X^q]$ in the case of fully ionized defect states.

**Table 5.** Formation energy $E^f(X^q)$ and thermodynamic transition levels $\epsilon(q/q')$ of $Zn_i$, $H_i$ and $H_o$ in ZnO calculated with DMC with the T-moves approach and a 32 atom supercell under Zn-rich conditions and $E_v$ = VBM, all values in eV. To evaluate $E^f(H_o)$, Zn-rich conditions were simulated employing the experimental enthalpy of formation of ZnO. Finite-size corrections due to image interactions were approximated within LDA.[a] The DMC statistical error is included in parenthesis.

| Defect | $E^f(X^0)$ | $E^f(X^{1+})$ | $E^f(X^{2+})$ | $\epsilon(2+/1+)$ | $\epsilon(2+/0)$ | $\epsilon(1+/0)$ |
|---|---|---|---|---|---|---|
| $Zn_i$ | 3.73(9) | 0.09(11) | -3.70(13) | 3.80(18) | 3.72(16) | 3.65(11) |
| $H_i$ | 1.89(9) | -1.61(11) | | | | 3.50(11) |
| $H_o$ | 1.99(8) | -1.78(10) | | | | 3.76(11) |
| $H_{2,int}$ | 2.33(9) | | | | | |

[a]The correction term $E_{corr}(X^q)$ applied to the DMC formation energy of $Zn_i^0$, $Zn_i^{1+}$, $Zn_i^{2+}$, $H_i^0$, $H_i^{1+}$, $H_o^0$, and $H_o^{1+}$ are -2.22, -0.60, 1.06, -1.56, 0.29, -1.24 and 0.11 eV, respectively.

Under *n*-type and Zn-rich conditions (**Figure 6**), DMC and HSE38 yield a similar relative stability order for the defects we have explored, i.e., $E^f(V_o) > E^f(H_o^{1+}) > E^f(H_i^{1-}) > E^f(H_{2,int}) > E^f(Zn_i^{2-})$. However, the formation energies evaluated with these two methods are different. The formation energies are ~0.5 eV higher when evaluated with DMC for most defects. The exception is $E^f(Zn_i^{2+})$, where DMC predicts a ~1 eV lower formation energy. The concentration of zinc interstitials in ZnO will be very low under *n*-type conditions as the formation energy is above 3 eV. Similarly, the formation energy for $H_i^{1+}$ and $H_O^{1+}$ found in DMC are over 1.4 eV, indicating that the concentration of these defects in ZnO will be low under *n*-type and Zn- and H-rich conditions. Under *p*-type conditions, $Zn_i^{2+}$ has the lowest formation energy according to our DMC calculations. $V_O^{2+}$ is ~1 eV higher in energy than $Zn_i^{2+}$ under these conditions. These results clearly contract with HSE calculations, where $V_O^{2+}$ is found to be ~0.6 eV more stable than $Zn_i^{2+}$ under *p*-type conditions. Nevertheless, the formation energies of these defects are below -2.5 eV under *p*-type and Zn-rich conditions. Therefore, a strong compensation of holes is expected under these conditions.[92,99]

IV. **CONCLUSIONS**

In summary, we have applied the fixed-phase DMC method, which is a fully *ab-initio* many-body approach, to study the structural stability and the energetics of intrinsic defects and hydrogen impurities in ZnO. The study shows that DMC is now a practical method that can be used to accurately characterize multiple properties of materials that are challenging for density functional theory approximations. Moreover, a comparison of results from DMC and HSE hybrid functionals shows that these DFT approximations in general yield a correct qualitative description of these materials. For instance, both methods show similar trends for phase stability and defect energetics under *n*-type conditions. However, the formation energy of defects in ZnO can differ by over 0.5 eV when evaluated with DMC and HSE functionals. Moreover, we found that parameterization of hybrid functionals to reproduce a particular property in transition-metal oxides (e.g., the band-gap) is not enough for a quantitative description of other properties, such as the energetics of the oxygen vacancy in ZnO. Finally, we note that there are several outstanding challenges (such as the best choice of pseudopotentials, magnitude of the fixed-node errors and finite size effects) that remain to be overcome for highly accurate *ab*-initio calculations of defects in ZnO. For a defect like the oxygen vacancy in ZnO, our calculations indicate that these effects are less significant than the differences we found between the DMC and DFT results. However, for delocalized defects like zinc interstitial and hydrogen impurities in ZnO, the finite size effects remain a major challenge.

ACKNOWLEDGMENT

We thank H. Dixit and L. Shulenburger for providing access to pseudopotential datasets and A. Zunger for helpful discussions and pointing us to the measurements in Refs. 107 and 108. The work was supported by the Materials Sciences & Engineering Division of the Office of Basic Energy Sciences, U.S. Department of Energy. Paul R. C. Kent was supported by the Scientific User Facilities Division, Office of Basic Energy Sciences, U.S. Department of Energy. Computational resources were provided by the Oak Ridge Leadership Computing Facility at the Oak Ridge National Laboratory, supported by the Office of Science of the U.S. Department of Energy under Contract No. DE-AC05-00OR22725.

# Supplemental Material for

# Structural Stability and Defect Energetics of ZnO from Diffusion Quantum Monte Carlo

Juan A. Santana,[1] Jaron T. Krogel,[1] Jeongnim Kim,[1] Paul R. C. Kent,[2, 3] Fernando A. Reboredo[1, a]

[1] Materials Science and Technology Division, Oak Ridge National Laboratory, Oak Ridge, Tennessee 37831, USA

[2] Center for Nanophase Materials Sciences, Oak Ridge National Laboratory, Oak Ridge, Tennessee 37831, USA

[3] Computer Science and Mathematics Division, Oak Ridge National Laboratory, Oak Ridge, Tennessee 37831, USA

## I. ZN PSEUDOPOTENTIAL

The Ionization Potential (IP) of the Zn atom evaluated with diffusion quantum Monte Carlo (DMC) and our pseudopotential (PP) deviates from the experiment by 0.14 eV. This deviation can come from the incomplete treatment of relativistic effects, the fixed-node approximation, the nonlocal PP approximation in DMC and missing core-valence electron correlation from the PP approximation.[1] Relativistic effects on the IP can be estimated from fully relativistic and non-relativistic calculations. The IP of the Zn atom (relative to the experimental value) evaluated with the all-electron fully relativistic multireference perturbation theory (MRMP)[2] method and DMC with our PP are shown in **Figure 1**. MRMP at the relativistic limit reproduces the experimental IP of Zn to within 0.03 eV. Relativistic effects on the IP of Zn account for 0.18 eV in MRMP, very similar to the value in DMC with PP, 0.17 eV. In LDA, the IP of Zn evaluated with PPs with and without relativistic corrections differs by 0.20 eV. The missing relativistic effects in the PP, mainly spin-orbit coupling, mostly cancel out for the IP.[1]

Fixed-node errors can be partially explored[3,4] by generating single-particle orbitals with hybrid and Hubbard-corrected functionals. As shown in **Figure 1**, the IP of Zn evaluated with DMC and orbitals from LDA and the Heyd-Scuseria-Ernzerhof (HSE)[5] hybrid functional are statistically equal. Effects of the approximations to evaluate nonlocal PP in DMC can be explored by comparing the IP evaluated with DMC within the locality approximation[6] and the T-move approach.[7] The IP evaluated with the T-moves method deviated from the experiment by 0.12 eV more than the value from the locality approximation. The effect of the PP approximation can be explored by comparing all-electron (AE) and PP calculations. We performed DMC-AE employing single-particle orbitals from Hartree-Fock (HF). The deviation of the IP of Zn evaluated this DMC-AE from the experiment is -0.483(13) eV (data not shown). This deviation is much larger than missing relativistic effects in HF (0.18 eV), indicating that the HF single-particle orbitals defining the nodal-surface are poor; the results cannot be directly compared with the corresponding DMC-PP calculations. Alternatively, we can examine the IP of Zn evaluated within LDA-AE and LDA-PP instead of DMC-AE and DMC-PP. The IP of Zn evaluated[8] with LDA-AE and LDA-PP is 10.17 and 10.23 eV, respectively; LDA-AE is 0.05 eV closer to the experiment. These test calculations indicate that the error of 0.14 eV for the IP of Zn calculated with DMC-PP comes mainly from nodal errors and the PP approximation.

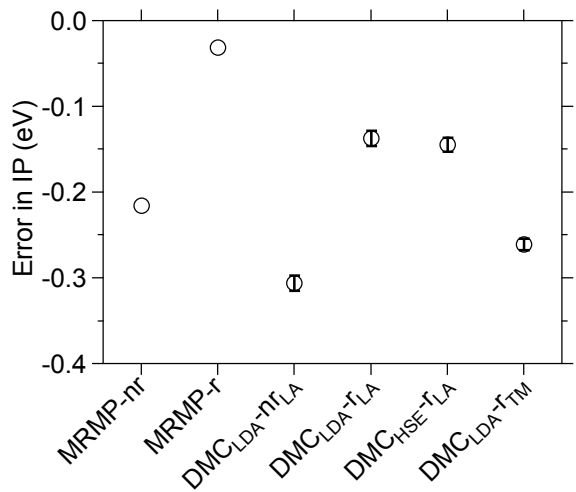

**Figure 1**. Theory-experiment deviation for the ionization potential of the Zn atom. Calculations were performed at the non-relativistic (nr) and relativistic (r) limits with the multireference perturbation theory (MRMP) and DMC methods. DMC calculations were performed within the locality approximation (LA) and the T-moves (TM) approach and with single-particle orbitals generated with LDA and HSE. Standard deviation error bars show statistical uncertainties in the DMC data.

## II. TIME STEP AND FINITE SIZE ERRORS

### A. Zn crystal (hcp)

To evaluate the structural parameters of the Zn crystal within DMC, we first estimated time step and finite-size (FS) errors. These test calculations were performed with the hexagonal close-packed structure of Zn with the experimental lattice parameters.[9] Time-step errors were studied by calculating the cohesive energy of Zn with DMC at three-time steps (0.02, 0.01 and 0.005 Ha$^{-1}$). Calculations were performed with a 2×2×1 (8 atom) supercell and twist boundary conditions on a 4×4×4 grid; results are shown in the upper panel of **Figure 2**. These test calculations were performed within the locality approximation and the T-moves approach. As a compromise between simulation time and computational cost, calculations of the cohesive energy of Zn crystal at different volumes were performed within the locality approximation with a time step of 0.01 Ha$^{-1}$. The DMC cohesive energies evaluated with this time-step were corrected by 0.20(1) eV/formula unit (f.u.) to extrapolate to time-step 0.

To estimate FS errors, we explored one-body FS errors[10] by evaluating the DMC energy of a 2×2×1 supercell and twisted boundary conditions[11] on 8×8×8, 6×6×4 and 4×4×4 grids. The DMC energy with 512, 144, and 64 twists were similar to within 0.01 eV/f.u. To study two-body FS errors,[10] we performed DMC calculation with 2×2×1, 2×2×2, and 3×3×2 supercells with twisted boundary conditions on a 6×6×4, 6×6×2 and 4×4×2 grid, respectively; results are shown in the central panel of **Figure 2**. Two-body FS errors can be effectively removed by employing the method of Ref. 12, or the model periodic Coulomb (MPC) interaction[10,13] corrected for kinetic contributions.[10,13] For Zn crystal, there are not significant residual FS errors after using either of these methods. We employed the method of Ref. 12 in our calculations at different volumes because this method shows only a weak supercell-shape dependence;[12] for instance, note that the DMC energy for the 2×2×2 supercell ($1/N = 1/16$) corrected with MPC in **Figure 2** deviates from the more converged results obtained with the method of Ref. 12.

We performed DMC calculations at different volumes to evaluate the equilibrium properties of Zn crystal. Calculations were carried out with the 2×2×1 supercell and twisted boundary conditions[11] on a 4×4×4 grid. The DMC energy of Zn crystal is plotted as a function of volume in the lower panel of **Figure 2**.

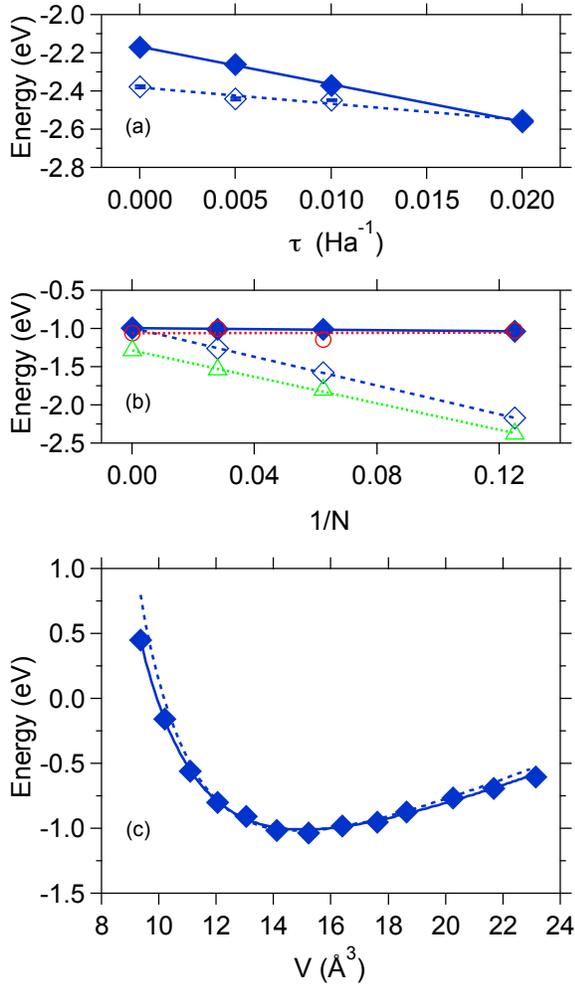

**Figure 2.** (a) DMC energy of Zn crystal evaluated within the locality approximation (full diamonds) and the T-moves approach (empty diamonds) as a function of imaginary time step (extrapolated energies are included at time step 0). (b) DMC energy (empty diamonds; locality approximation, empty triangles; T-moves approach) and DMC energy (within the locality approximation) corrected for 2-body FS errors with the method of Ref. 12 (full diamonds) and MPC[10,13] (empty circles) as a function of supercell size (cell of 1/36, 1/16, and 1/8 atoms); extrapolated energies are included at $1/N = 0$. (c) DMC energy versus volume per formula unit together with fitted Murnaghan (solid curve) and Vinet (dashed curve) EOS. In (b), note the overlapping values at $1/N = 1/8$ and $1/36$. The DMC energies in (b) and (c) have been corrected for time step errors as estimated in (a). Energies are per formula unit and relative to the Zn atom. The statistical uncertainty in DMC is smaller than the symbol size.

## B. ZnO crystals

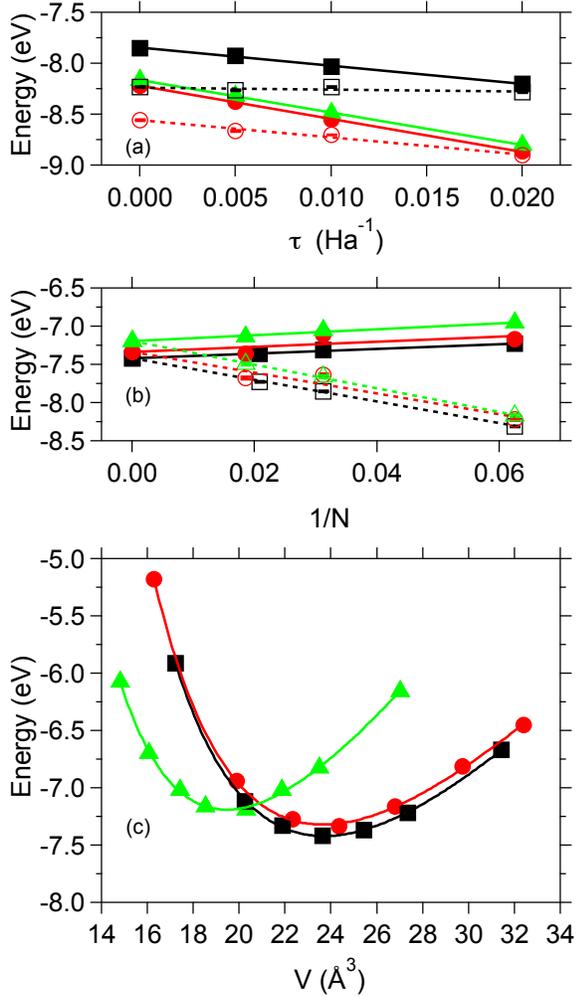

**Figure 3.** (a) DMC energies of ZnO in the rock salt (triangles), zinc blende (circles) and wurtzite (squares) phases evaluated within the locality approximation (full symbols) and the T-moves approach (empty symbols) as a function of imaginary time step (extrapolated energies are included at time step 0). (b) DMC energy (empty symbols), and DMC energy corrected for 2-body FS errors with the method of Ref. 12 (full symbols) as a function of supercell size; extrapolated energies are included at $1/N = 0$). (c) DMC energy versus volume per formula unit together with fitted Murnaghan EOS. The DMC energies in (b) and (c) have been corrected for time step errors as determined in (a) within the locality approximation. In (c), residual FS errors were corrected as estimated in (b). Energies are per formula unit and relative to the Zn and O atoms. The statistical uncertainty in DMC is smaller than the symbol size.

Similar to the Zn crystal, we first estimated time step and FS errors to calculate the structural properties of ZnO with DMC. Time step errors were studied by calculating the cohesive energy of ZnO with DMC at three-time steps (0.02, 0.01 and 0.005 Ha$^{-1}$). Calculations were performed with 2×2×2 (16 atom) supercells of ZnO in the B1, B3 and B4 phases. For the B3 and B4 phases, calculations were performed within the locality approximation and the T-moves approach; results are shown in the upper panel of **Figure 3**. Further calculations of the cohesive energy of the ZnO phases at different volumes were performed with a time step of 0.01 Ha$^{-1}$. The DMC cohesive energies of the B1, B3, and B4 phases were corrected for time-step error by 0.32(1), 0.33(1) and 0.181(8) eV/f.u, respectively.

One-body FS errors were accounted for by using twisted boundary conditions[11] on more than 144 twists for each phase. This number of twists was enough to converge the DMC energy to within 0.01 eV/f.u. Two-body FS errors[10] were studied with supercell of 16, 32 and 54 atoms for the B1 and B3 phases and 16, 32 and 48 for the B4 phase; results are shown in the central panel of **Figure 3**. The bulk of the two-body FS errors is removed by employing the method of Ref. 12 There are residual FS errors after correcting for the bulk of two-body FS errors. The residual error was obtained by extrapolating the partially corrected energies as a function of 1/N to the thermodynamic limit. To account for residual FS errors, the cohesive energy evaluated at different volumes was corrected by 0.24(2) and 0.17(2) and 0.11 (2) eV/f.u for the B1, B3 and B4 phases, respectively. Variation of the residual FS errors due to change in volume was not considered.

The structural parameters of ZnO were determined evaluating the DMC cohesive energy at different volumes with 16 atom supercells for the B1 and B3 phases and a 32 atom supercell for the B4 phase. The DMC energy of ZnO in the B1, B3, and B4 phases is plotted as a function of volume in the lower panel of **Figure 3**.

## III. DEFECTS – $V_O$ TIME STEP ERROR

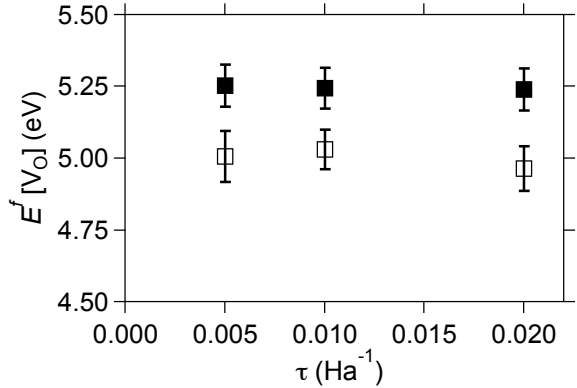

**Figure 4**. Formation energy of the neutral oxygen vacancy in ZnO evaluated with DMC within the locality approximation (empty squares) and the T-moves approach (full squares) as a function of the DMC time step. Calculations were performed with a 32 atom supercell and twisted boundary conditions on a 4×4×2 grid (under O-rich conditions). Standard deviation error bars show statistical uncertainties in the DMC data.

As shown in **Figure 4**, time step errors mostly cancel out for the formation energy of the oxygen vacancy.